\documentclass[prb,preprint]{revtex4}
\usepackage{amsmath}
\usepackage{amsfonts}
\usepackage{graphicx}

\begin{document}

\title{
Interpretation of Diffusing Wave Spectra in Nontrivial Systems
}

\author{George D. J. Phillies}
\email{phillies@wpi.edu}

\affiliation{Department of Physics, Worcester Polytechnic
Institute,Worcester, MA 01609}

\begin{abstract}

Mathematical methods previously used (Phillies, J. Chem.\ Phys., {\bf 122} 224905 (2005)) to interpret quasielastic light scattering spectroscopy (QELSS) spectra are here applied to relate diffusing wave spectroscopy (DWS) spectra to the moments $\overline{X^{2n}}$ of particle displacements in the solution under study.  DWS spectra of optical probes are like QELSS spectra in that in general they are not determined solely by the second moment $\overline{X^{2}}$.  In each case, the relationship between the spectrum and the particle motions arises from the field correlation function $g^{(1)}_{s}(t)$ for a single quasi-elastic scattering event. In most physically interesting cases, $g^{(1)}_{s}(t)$ receives except at the shortest times large contributions from higher moments $\overline{X(t)^{2n}}$, $n >1$.  As has long been known, the idealized form $g^{(1)}_{s}(t) =\exp(-2 q^{2} \overline{X(t)^{2}})$, sometimes invoked to interpret DWS and QELSS spectra, only refers to (adequately) monodisperse, noninteracting, probes in purely Newtonian liquids and is erroneous for polydisperse particles, interacting particles, or particles in viscoelastic complex fluids.  Furthermore, in DWS experiments fluctuations (for multiple scattering paths of fixed length) in the number of scattering events and the total-square scattering vector significantly modify the spectrum.    

\end{abstract}

\maketitle

\section{Introduction}

Three decades ago, Hallett and students\cite{hallett1974a,hallett1976a} used quasielastic light scattering spectroscopy (QELSS) to study dilute, intensely scattering probe spheres diffusing through polymer solutions.  The spheres were used as optical probes of the polymer solution's rheology.  By employing the Stokes-Einstein equation
\begin{equation}
    D_{p} = \frac{k_{B} T}{6 \pi \eta_{\mu} R},
    \label{eq:seeq}
\end{equation}
an apparent (micro)viscosity $\eta_{\mu}$ was inferred from the diffusion coefficient $D_{p}$.  Here $k_{B}$, $T$, and $R$ are Boltzmann's constant, the absolute temperature, and the probe radius, respectively.  Hallet, et al.'s $\eta_{\mu}$ were nearly equal to the viscosities $\eta$ determined with classical instruments.  These experiments were a logical continuation of earlier work of Laurent, et al.\cite{laurent1963a}, who used ultracentrifugation to study the sedimentation of mesoscopic colloids through various complex fluids.  

QELSS, fluorescence recovery after photobleaching, and related techniques have since been used to study the diffusion of optical probes through a wide variety of complex fluids, such as polymer\cite{russo1993a,lin1982a}, surfactant\cite{phillies1993a}, colloid\cite{pusey1982a}, and protein\cite{ullmann1985b} solutions.  These studies are united under the cognomen {\em Optical Probe Diffusion} (OPD).  Effective ways to control the binding of at least some polymers to some probes\cite{russo1989a} are known, permitting calibrated measurements of the microviscosity and the measurement of the thickness of the bound polymer\cite{ullmann1983a}.  It was early recognized that the microviscosity and the viscosity $\eta$ obtained from capillary viscometers are often not the same.  In a few cases, exotic reentrant behavior is observed\cite{ullmann1985a}, in which $\eta_{\mu} \neq \eta$, but only over a narrow band of concentrations.  Systematic studies of $\eta_{\mu}$, as a function of the concentration and molecular weight of matrix polymers in solution\cite{yang1988a,furukawa1991a}, investigated whether optical probe diffusion is only an expensive replacement (albeit helpful with microscale samples) for an inexpensive capillary viscometer, or whether it gives interesting physical results.  Fortunately, the latter case is correct: Optical Probe Diffusion gives novel information on polymer dynamics.

 Artifacts such as probe aggregation or polymer adsorption by the probe cause the probe to move too slowly, leading to $\eta_{\mu} > \eta$, thus permitting a ready separation of several important classes of artifact from physically interesting results.  In the absence of artifacts, it is generally the case that $\eta_{\mu}=\eta$ or that $\eta_{\mu} < \eta$ (sometimes $\eta_{\mu} \ll \eta$). It is incorrect to propose that $\eta > \eta_{\mu}$ could arise from the non-zero shear rate of capillary viscometers. For $\eta > \eta_{\mu}$ to arise from viscometer shear, the polymer solution would need to be shear {\em thickening}.  Shear thickening is a rare phenomenon not encountered with most polymers that have been studied with probe diffusion.

QELSS measures the field correlation function $g^{(1)}_{s}(q,t)$ of the light single-scattered by the probes.  QELSS spectra have recently sometimes been interpreted under the cognomen {\em microrheology}, in which it is claimed that QELSS spectra are related to particle displacements via 
\begin{equation}
      g^{(1)}_{s}(q,t) = \exp( - q^{2} \overline{X(t)^{2}}/2),
      \label{eq:g1qelssform}
\end{equation}
$q$ being the scattering vector determined by the laser wavelength and apparatus geometry.  In microrheology, from $\overline{X(t)^{2}}$, a time-dependent diffusion coefficient is inferred from the supposed form $\overline{X(t)^{2}} = 2 t D(t)$.  Finally, a time- or frequency-dependent generalization of the Stokes-Einstein equation and time-frequency transforms are used to infer from $D(t)$ a frequency-dependent viscosity, or storage and loss moduli, corresponding to the original DWS spectrum.  This process for determining $G'(\omega)$ and  $G''(\omega)$ from $\overline{X(t)^{2}}$ may be compared with determinations using classical mechanical means.

Equations superficially similar to eq \ref{eq:g1qelssform} appear in Berne and Pecora\cite{berne1976a}.  However, as is made clear in Berne and Pecora, equation \ref{eq:g1qelssform} only refers to the diffusion of identical, true Brownian particles whose motion is governed by the Langevin model.  In this model, there are no memory effects such as those arising from solvent viscoelasticity.  For Brownian particles that satisfy the Langevin model, $\overline{X(t)^{2}} = 2 D t$, where $D$ is the {\em time-independent} diffusion \emph{constant}. In the QELSS literature, eq \ref{eq:g1qelssform} is only applied to certain simple cases, e.g., monodisperse polystyrene spheres in water. 

The motions of diffusing particles in viscoelastic fluids do not follow the Langevin equation, because the random forces have non-zero correlation times.  Correspondingly, it is incorrect to apply eq \ref{eq:g1qelssform} to interpret QELSS spectra of probes in viscoelastic fluids.  Equation \ref{eq:g1qelssform} also incorrectly predicts $\log (g^{(1)}_{s}(q,t)) \sim q^{2}$  as a uniform result.  Modifications to the Langevin equation for physically realistic, i.e., viscoelastic at adequately high frequency, motions are encompassed by the Mori formalism\cite{mori}, which moves from the fundamental, microscopic Liouville equation for particle motion to a reduced-variable description in which the random force and corresponding drag coefficient are described by memory functions.

In order to clarify the interpretation of QELSS spectra, we recently examined\cite{phillies2005a} the relationship between $g^{(1)}_{s}(q,t)$ and the mean-$n^{\rm th}$ moments $\overline{X(t)^{n}}$ of the particle displacement along a single coordinate axis.  The higher-order moments incorporate fluctuations in the particle displacements around their mean-square values. We\cite{phillies2005a} showed: The odd moments of $X$ vanish.  In general, all even moments $\overline{X(t)^{n}}$ of the displacement contribute to  $g^{(1)}_{s}(q,t)$. A simple diagnostic identifies the special case in which eq \ref{eq:g1qelssform} is correct:  If the QELSS spectrum agrees with eq \ref{eq:g1qelssform}, then from Doob's First Theorem\cite{doob1942a} the QELSS spectrum is \emph{necessarily} a pure exponential with a time-independent diffusion constant.  Doob's theorem is a \emph{purely mathematical} result on zero-correlation-time random walks, and uses no physical arguments or rationales.  Conversely, {\em if the light scattering spectrum is not a single exponential, then the contrapositive form of Doob's theorem shows that eq \ref{eq:g1qelssform} cannot describe the QELSS spectrum};  $g^{(1)}_{s}(q,t)$ contains non-trivial contributions from all higher-order even moments $\overline{X(t)^{2n}}$.  

Equation \ref{eq:g1qelssform} describes the motion of a solution of Brownian particles in an ideal simple fluid.  If the particles are not monodisperse, or if the fluid is viscoelastic, the spectrum ceases to be a simple exponential, and $g^{(1)}_{s}(t)$ does not follow eq.\ \ref{eq:g1qelssform}.  It is not possible to mask the contributions of the higher moments of $X$ by going to small $q$.  While eq \ref{eq:g1qelssform} cannot be used to interpret light scattering spectra of probe particles in most complex fluids, nothing is wrong with the underlying QELSS spectra that eq \ref{eq:g1qelssform} has been used to analyze.  Reanalysis of those spectra using conventional means such as cumulants or mode decomposition may well lead to useful information.

Recently, an alternative technique that uses light scattering to study the diffusion of optical probes has been proposed\cite{maret1987a,weitz1993a}.  Diffusing Wave Spectroscopy (DWS) studies light that has repeatedly been scattered by an intensely-scattering, colorless, opaque complex fluid. In DWS experiments considered here, scattering is due to diffusing mesoscopic particles that lead to multiple scattering of the incident light.  The individual scattering events are describable as quasi-elastic single scattering; scattering events are far enough apart from each other that the motions of particles in a multiple scattering series are uncorrelated.  

The underlying experimental apparatuses in QELSS and DWS are the same, with a laser, scattering sample, and photomultiplier tube or equivalent photodetector.  Differences appear in the scattering cell, probe concentration, and preferred scattering angles.  In DWS and QELSS, the intensity autocorrelation function is obtained with a conventional digital autocorrelator; the time scales instrumentally accessible to QELSS and DWS are determined by the same instruments and are the same.  The fluids and probes used by DWS are the same as those studied by Hallett, et al.\cite{hallett1974a,hallett1976a} and many successors. OPD and DWS differ secondarily in the opacity of their samples and primarily in the assumptions and formalism used to interpret their spectra.  The cognomen Diffusing Wave Spectroscopy refers to a particular model for interpreting the field correlation function of the multiple-scattered light. In standard treatments of DWS in diffusing systems, it is proposed that DWS determines the mean-square diffusive displacement $\overline{X(t)^{2}}$ of individual particles during time interval $t$.   DWS has also been applied to nondiffusive systems as described below.

This paper treats the contribution of particle motion and scattering path fluctuations to DWS spectra. It is shown below that fluctuations $\overline{X^{2n}}$, $n >1$, in the particle displacement in principle make non-negligible contributions to DWS spectra.  Furthermore, even for paths of fixed length, these fluctuations are mixed with fluctuations from path to path in the number of scattering events and in the total-square scattering vector.  Fluctuations lead to interesting challenges for the interpretation of DWS spectra in terms of particle motion. 

\section{Physical Basis of the Diffusing Wave Spectrum}

We first examine general considerations\cite{pine1990a,weitz1993a,mackintosh1989a} that permit calculation of the diffusing wave spectrum.  In essence, in a DWS experiment the light is scattered along many different paths before it reaches the detector.  The scattered field incident on the photodetector is the amplitude-weighted coherent sum of the electric fields of the light travelling over all multiply-scattered paths.  To describe the light scattered along a single path $P$ having $N$ scattering events, we take the positions of the $N$ scattering particles to be ${\bf r}_{i}(t)$ for $i \in (1,N)$.  The source is located at ${\bf r}_{0}$; the detector is located at  ${\bf r}_{N+1}$.  Unlike the particle positions, the source and detector positions are independent of time.  The light initially has wavevector ${\bf k_{0}}$; the light scattered from particle $i$ to particle $i+1$ has wavevector ${\bf k}_{i}(t)$.  The light emerges from the system and proceeds to the detector with wavevector ${\bf k}_{N}$.  The wavevectors ${\bf k_{0}}$ and ${\bf k_{N}}$ are the same for every scattering path and at all times. The ${\bf k}_{i}$ for $1 \leq i < N$ are functions of time because they link pairs of particles, and the particles move.  Paths differ in the number of scattering particles and the ordered list of particles by which the light was scattered.  For the scattering along a specific path $P$ the phase shift is
\begin{equation}
    \Delta_{P} \phi(t) = - {\bf k_{0}}(t)\cdot {\bf r}_{0} - \sum_{i=1}^{N} {\bf r}_{i}(t) \cdot ( \ {\bf k_{i}}(t) - {\bf k_{i-1}}(t)\ ) + {\bf r}_{N+1}(t) \cdot {\bf k_{N}}(t)
    \label{eq:phaseshift}
\end{equation}
and the total field scattered along the available paths is
\begin{equation}
       E(t) = \sum E_{P} \exp (i \Delta_{P} \phi(t)).
        \label{eq:allpaths}
\end{equation}
The summation is taken over the list of all allowed paths. The list of all allowed paths is the set of all ordered lists, of any length greater than zero, of particles in the system, subject to the constraint that no two adjoining elements in a list may refer to the same particle.  $E_{P}$ is the scattering amplitude associated with a particular path $P$.  The phase shift $\Delta_{P} \phi(t)$ and scattering amplitude depend on time because the particles move.  Because of translational invariance, the distribution of $\exp (i \Delta_{P} \phi(t))$ is flat modulo $2 \pi$, and the distribution of $E_{P}$ is well-behaved. $E(t)$ is therefore the sum of a large number of independent nearly-identically distributed random variables, so it is reasonable to infer from the central limit theorem that $E(t)$ has a gaussian distribution.  Because the $E(t)$ have joint Gaussian distributions, the intensity autocorrelation function $S(t)$ is determined by the field correlation function of the multiply scattered light.

This description of multiple scattering reduces to the standard description of single-scattered light if the number of particles along a path is constrained to $N=1$.  By comparison with that description, the ${\bf r}_{i}(t)$ are the optical centers of mass of the diffusing particles, the single-particle structure factors arising from internal interference within each particle being contained in $E_{P}$.  The standard description of scattering in terms of particle positions continues to be correct if the particles are non-dilute, and is the basis for calculating QELSS spectra of concentrated particle suspensions.

The field correlation function depends on the phase shift as
\begin{equation}
     g^{(1)}(t) \equiv \langle E(0) E^{*}(t) \rangle = \left\langle \sum_{P,P'} E_{P}(0) E^{*}_{P'}(t) \exp \left[i \Delta_{P} \phi(t) -     i \Delta_{P'} \phi(0)\right]\right\rangle.
     \label{eq:fieldcorrelation}     
\end{equation}
Because the paths are independent, in the double sum on $P$ and $P'$ only terms with $P=P'$ are on the average non-vanishing.  For those terms, the change between $0$ and $t$ in the phase is rewritten by applying: the definition eq \ref{eq:phaseshift} of the phase shift; the requirement that the source and detector locations and the initial and final wavevectors are independent of time; the definition ${\bf q}_{i} (t) = {\bf k}_{i}(t) - {\bf k}_{i-1}(t)$ of the scattering vector; and addition of $0$ in the form ${\bf q}_{i} (0) \cdot {\bf r}_{i}(t)-{\bf q}_{i} (0) \cdot {\bf r}_{i}(t)$. Re-ordering terms gives 
\begin{equation}
     g^{(1)}(t) = \left\langle \sum_{P}  E_{P}(0) E^{*}_{P}(t)  \exp \left(- i \sum_{i=1}^{N} {\bf q}_{i} (0) \cdot \delta {\bf r}_{i}(t)  - \delta {\bf q}_{i}(t) \cdot {\bf r}_{i}(t) \right)\right\rangle,
     \label{eq:fieldcorrelation2}     
\end{equation}
where $\delta {\bf r}_{i}(t) = {\bf r}_{i}(t) - {\bf r}_{i}(0)$ is the particle displacement, and $\delta {\bf q}_{i}(t) = {\bf q}_{i} (t)- {\bf q}_{i} (0) $ is the change in scattering vector $i$ between times $0$ and $t$.  The distance between scattering events is generally very large compared to the distances over which particles move before $g^{(1)}(t)$ has relaxed to zero, and the $\mid {\bf k}_{i}\mid$ always have the same magnitude, so ${\bf q}_{i}$ can only be altered by changing the angle between $ {\bf k}_{i-1}$ and $ {\bf k}_{i}$.  In consequence, it is asserted that $\delta {\bf q}_{i} (t)$ term is extremely small relative to the ${\bf q}_{i} (0)$ term, because $\mid \delta {\bf q}_{i} (t) \mid / \mid {\bf q}_{i} (0)\mid $ and $\mid \delta {\bf r}_{i}(t) \mid / \mid {\bf r}_{i}(t) \mid$ are of the same order. The scattering amplitudes within the $E_{P}$ only change as the scattering angles in each scattering event change, so, by the same rationale, for each path  $E_{P}(t)$ is very nearly independent of time over the times of interest.  Furthermore, while the total of the scattering vectors must match the initial and final wave vector, if the number of scattering events along a path is large the constraint on the total scattering vector has very little effect on the intermediate scattering vectors.

Under these approximations, the field correlation function reduces to
\begin{equation}
     g^{(1)}(t) = \left \langle \sum_{P} \mid E_{P}(0) \mid^{2} \exp \left(- i \sum_{i=1}^{N} {\bf q}_{i} (0) \cdot \delta {\bf r}_{i}(t)  \right) \right\rangle,
      \label{eq:g1explicit}
\end{equation}
as shown by Weitz and Pine\cite{weitz1993a}.  The outermost average $\langle \cdots \rangle$ extends over all particle positions and subsequent displacements. No assumption has thusfar been made about interactions between scattering particles.

In order to evaluate this form, Weitz and Pine\cite{weitz1993a} impose three approximations.  Essentially equivalent approximations are imposed by other references\cite{mackintosh1989a,pine1990a}.  Each approximation replaces a fluctuating quantity with an average value.  The three approximations are:

Approximation 1) The exponential over the sum of particle displacements can be factorized, namely
\begin{equation}
\sum_{P} \left \langle  \mid E_{P}(0) \mid^{2} \exp \left(- i \sum_{i=1}^{N} {\bf q}_{i} (0) \cdot \delta {\bf r}_{i}(t)  \right) \right\rangle =  \sum_{P}   \mid E_{P}(0) \mid^{2}  \prod_{i} \left \langle\exp \left(- i {\bf q}_{i} (0) \cdot \delta {\bf r}_{i}(t)  \right) \right\rangle,
\label{eq:factorize}
\end{equation} 
so that averages over particle displacements can be taken separately over each particle.  The factorization is valid if the scattering points\emph{ along each path} are dilute, so that the particle displacements $\delta {\bf r}_{1}(t), \delta {\bf r}_{2}(t),\ldots$ are independent, because under this condition the distribution function $P^{(N)}(\delta {\bf r}_{1}(t), \delta {\bf r}_{2}(t),\ldots )$ for the simultaneous displacements of the $N$ particles of a path factors into a product of $N$ single-particle displacement distributions $P^{(1)}(\delta {\bf r}_{i}(t))$, one for each particle.  In representative DWS experiments, mean free paths for optical scattering are reported as hundreds of microns\cite{pine1990a}, while the effective range of the very-long-range interparticle hydrodynamic interactions is some modest multiple of the particle radius or a modest number of microns, so the typical distance between scattering events is indeed far larger than the range over which particles can influence each other's motions.

The assertion that the scattering points along each path are dilute does not imply that the scatterers are dilute. If the scattering cross-section of the scatterers is not too large, a given photon will be scattered by a particle, pass through many intervening particles, and finally be scattered by another, distant particle.  The scattering particles  may themselves be concentrated.  The physical requirement on dilution used in ref.\ \onlinecite{weitz1993a} is that the mean free path between serial scattering events of a given photon is much longer than the range of the interparticle interactions.  The motion of each scatterer in a path may very well be influenced by its near neighbors, but those near neighbors are almost never parts of the same path.  While two very nearby particles could be involved in one path, this possibility involves a small fraction of the entirety of paths. Indeed, if most paths $P$ included pairs of particles that were close enough to each other to interact with each other, the above description of DWS spectra would fail completely: Eq \ref{eq:g1explicit} would not follow from eq \ref{eq:fieldcorrelation2}, because the terms of eq \ref{eq:fieldcorrelation2} in $\delta {\bf q}_{i}(t)$ would become significant.

Finally, refs. \onlinecite{mackintosh1989a,pine1990a,weitz1993a} propose that the average over the individual particle displacements may be approximated as 
\begin{equation}
     \langle \exp \left(- i {\bf q}_{i} (0) \cdot \delta {\bf r}_{i}(t)  \right)\rangle \approx \exp\left(- {\bf q}_{i}^{2} \overline{X(t)^{2}}/2 \right).
     \label{eq:q2r2}
\end{equation}
where $\overline{X(t)^{2}} = \langle (\hat{q}_{i} \cdot \delta {\bf r}_{i}(t))^2 \rangle$. A variety of rationales for eq \ref{eq:q2r2} appear in the literature, as discussed below.  Pine, et al.\cite{pine1990a} show how eq \ref{eq:q2r2} can be replaced for nondiffusing particles when an alternative form is known \emph{a priori}. 

Approximation 2) Each scattering event has its own scattering angle and corresponding  scattering vector $\mid {\bf q}_{i}\mid$, which are approximated via $\langle \exp ( - {\bf q}_{i}^{2}  \overline{X^{2}}/2) \rangle \rightarrow  \exp(- \overline{q^{2}}\ \overline{X^{2}}/2)$, so that the scattering vector of each scattering event is replaced by a weighted average $\overline{q^{2}}$ of all scattering vectors.  Note that each pair ${\bf q}_{i}, {\bf q}_{i+1}$ of scattering vectors shares a common ${\bf k}_{i}$, so that $\langle {\bf q}_{i} \cdot {\bf q}_{i+1} \rangle \neq 0$.   So long as the particle displacements $\delta {\bf r}_{i}(t)$ and $\delta {\bf r}_{i+1}(t)$ are independent the cross-correlations in the scattering vectors do not affect the calculation.  

Approximation 3) The number $N$ of scattering events in a phase factor $\exp \left(- i \sum_{i=1}^{N} {\bf q}_{i} (0) \cdot \delta {\bf r}_{i}(t)  \right)$ is approximated as being entirely determined by the opacity of the medium and the length of each path, so that all paths of a given length have exactly the same number of scattering events. Weitz and Pine\cite{weitz1993a} propose that light propagation is effectively diffusive, there exists a mean distance $l^{*}$ over which the direction of light propagation decorrelates, the number of scattering events for a path of length $s$ is always exactly $N=s/\ell^{*}$, and the distribution of path lengths $P(s)$ between the entrance and exit windows of the scattering cell can be obtained from a diffusive first crossing problem.     

These approximations were\cite{weitz1993a} combined to predict that the field correlation function for DWS is
\begin{equation}
     g^{(1)}_{\rm DWS}(t) \ \propto \int_{0}^{\infty} ds \ P(s) \exp(- k_{0}^{2} \overline{X(t)^{2}} s/\ell^{*}),
    \label{eq:DWSspectrum}
\end{equation}
where $k_{0}$ is the wavevector of the original incident light and a mean scattering angle is linked by Ref.\ \onlinecite{weitz1993a} to $\ell^{*}$.  The model was solved for a suspension of identical particles performing simple Brownian motion, for which
\begin{equation}
    \overline{X(t)^{2}} = 2 D t.
    \label{eq:x2dt}
\end{equation}  For a light ray entering a cell at $x=0$, travelling diffusively through the cell to $x=L$, and emerging for the first time into the region $x \geq L$, ref.\ \onlinecite{weitz1993a} finds the distribution of path lengths, based on the diffusion equation with appropriate boundary conditions.  There are different solutions depending on whether one is uniformly illuminating the laser entrance window, is supplying a point source of light, or is supplying a narrow beam of light that has a Gaussian intensity profile.  For example, for uniform entrance window illumination eq 16.39a of ref \onlinecite{weitz1993a} gives for the DWS spectrum 
\begin{displaymath}
     g^{(1)}_{\rm DWS}(t) = \frac{L/\ell^{*} +4/3}{z_{o}/\ell^{*} + 2/3}
\left\{\sinh\left[\frac{z_{o}}{\ell^{*}} \left(\frac{6 t}{\tau}\right)^{0.5} \right]
+\frac{2}{3}\left(\frac{6 t}{\tau}\right)^{0.5}\cosh\left[\frac{z_{o}}{\ell^{*}} \left(\frac{6 t}{\tau}\right)^{0.5} \right]\right\} 
\end{displaymath}
\begin{equation}
\left\{\left(1+\frac{8t}{3\tau}\right)\sinh \left[\frac{L}{\ell^{*}}  \left(\frac{6 t}{\tau}\right)^{0.5} \right] + \frac{4}{3} \left(\frac{6 t}{\tau}\right)^{0.5}\cosh\left[\frac{L}{\ell^{*}} \left(\frac{6 t}{\tau}\right)^{0.5} \right]              \right\}^{-1}
    \label{eq:DWSspectrum2}
\end{equation}
where $z_{o} \approx \ell^{*}$ is the distance into the cell at which light motion has become diffusive, and 
\begin{equation}
      \tau = (D k_{o}^{2})^{-1}
      \label{eq:taudk0}
\end{equation} 
is a mean diffusion time.  

Even for an underlying simple exponential relaxation, the fluctuation in the total path length from path to path has caused the field correlation function for DWS to be quite complicated.  If eq \ref{eq:q2r2} were correct, then--ignoring the other approximations noted above--inversion of eq \ref{eq:DWSspectrum2} at a given $t$ could formally obtain from $g^{(1)}_{\rm DWS}(t)$ a value for $\tau$ at that $t$.  From $\tau$, eq \ref{eq:taudk0} formally gives a time-dependent $D$, whose frequency transform converts that time-dependent $D$ into the frequency-dependent storage and loss moduli.

\section{Fluctuation Corrections}

Each of the above three approximations replaced the average of a function of a quantity with the function of the average of that quantity.  Such replacements neglect the fluctuation in the quantity around its average, which is harmless if and only if the function is purely linear in the quantity being averaged.  However, $g^{(1)}_{DWS}(t)$ is an exponential of the fluctuating quantities.  It is inobvious that fluctuations in its arguments can be neglected. We first consider the separate fluctuations in $X$, $q^{2}$, and $N$ and then demonstrate their joint contribution to a DWS spectrum.
 
\subsection{Fluctuations in Particle Displacement}
 
To demonstrate the relationship between the single-scattering field correlation function
\begin{equation}
       g_{s}^{(1)}(q,t) = \langle \exp(- i {\bf q} \cdot {\bf r}_{i}(t) )\rangle
       \label{eq:g1qtaverage}
\end{equation}
and the mean particle displacements  $\overline{X(t)^{2n}}$, consider the Taylor series
\begin{equation}
     \langle \exp(- i {\bf q} \cdot {\bf r}_{i}(t) )\rangle = \left\langle \sum_{n=0}^{\infty} \frac{(- i {\bf q} \cdot {\bf r}_{i}(t))^{n} }{n!} \right\rangle.
     \label{eq:taylorseries}
\end{equation}
On the rhs, the average of the sum is the sum of the averages of the individual terms.  By reflection symmetry, averages over terms odd in ${\bf r}_{i}(t)$ vanish.  In the even terms, components of ${\bf r}_{i}$ that are orthogonal to ${\bf q}$ are killed by the scalar product.  Without loss of generality, the $x$-coordinate may locally be set so that the surviving component of the displacement lies along the $q$ axis, so  ${\bf q} \cdot {\bf r}_{i} = q x_{i}$, leading to
\begin{equation}
     \langle \exp(- i {\bf q} \cdot {\bf r}_{i}(t) )\rangle = \sum_{m=0}^{\infty} \frac{(-1)^m q^{2m} \overline{X^{2m}}}{(2m)!}
     \label{eq:displacementexpand}
\end{equation}
with $\langle x_{i}^{n} \rangle = \overline{X^{n}}$.  The series on the rhs of eq \ref{eq:displacementexpand} is the series
$\sum_{n=0}^{\infty} q^{2n} M_{n} /n!$ in $q^{2}$, with expansion coefficients 
\begin{equation}
     M_{o} = 1,
\end{equation}
\begin{equation}
     M_{1} = \frac{\overline{X^{2}}}{2},
\end{equation}
\begin{equation}
     M_{2} = \frac{\overline{X^{4}}}{12},
\end{equation}
\begin{equation} 
     M_{3} =\frac{\overline{X^{6}}}{120},
\end{equation}
etc.  

Equation \ref{eq:displacementexpand} may also be written as a cumulant expansion $\exp(\sum_{n} (- q^{2})^{n} K_{n}/n!)$, the $K_{n}$ being cumulants.  Expansion of the cumulant series as a power series in $q^{2}$ and comparison term by term shows 
\begin{equation}
 g^{(1)}_{s}(q,t) =
    \exp\left[- \left(q^2 \frac{\overline{X^{2}}}{2} - q^{4}\frac{
    (\overline{X^{4}} - 3
    \overline{X^{2}}^{2}) 
    }{24}+     q^{6} \frac{( 30 \overline{X^{2}}^{3} - 15   \overline{X^{2}} \  \overline{X^{4}} + \overline{X^{6}} )}{720} -\ldots \right) \right]
      \label{eq:displacementexpand2}
\end{equation}
with $K_{1} =\frac{\overline{X^{2}}}{2}$, $K_{2} = \frac{(\overline{X^{4}} - 3 \overline{X^{2}}^{2})  }{12}$, etc., the $K_{n}$ and $\overline{X^{2n}}$ being time-dependent.  The cumulants $K_{n}$ here differ from the cumulants in Section D, below.  If the distribution function for $X(t)$ were a Gaussian, then all cumulants above the first would vanish (e.g., $K_{2} = 0$), and eq \ref{eq:displacementexpand2} would reduce to eqs \ref{eq:g1qelssform} and \ref{eq:q2r2}.  However, $X(t)$ almost never has a Gaussian distribution in real systems.
   
\subsection{Fluctuations in the Scattering Vector}

For a single scattering event in which the light is deflected through an angle $\theta$, the magnitude of the scattering vector is
\begin{equation}
     q = 2 k_{0} \sin(\theta/2),
     \label{eq:qdefinition}
\end{equation}
where $k_{0} = 2 \pi n/\lambda$, with $n$ the index of refraction and $\lambda$ the light wavelength {\em in vacuo}. Cumulant expansions of spectra depend on powers of $q^{2}$ as
\begin{equation}
    q^{2} = 2 k_{0}^{2} (1 - \cos(\theta) ).
    \label{eq:q1definition}
\end{equation}
For DWS, the $q^{2}$ at the scattering points are independent from each other.  All scattering angles are permitted. The average over all scattering angles comes from an intensity-weighted average over all scattering directions.  For larger particles, the scattering is weighted by a particle form factor but always encompasses a non-zero range of angles.   For small particles, no $\theta$ is preferred. However, scattering from small particles is not isotropic, because light is a vector field.  Scattering from small particles is described by dipole radiation, whose amplitude is proportional to $\sin(\psi)$, $\psi$ being the angle between the direction of the scattered light and the direction of the polarization of the incident light. 

While it is true that light emerging from a turbid medium is depolarized, turbidity depolarization reflects the presence of many different paths, each of which rotates incident linearly polarized light through a different angle.  If linearly polarized light is scattered by a typical small particle or a dielectric sphere (the typical probe) of any size, the scattered light from that one scattering event remains linearly polarized, though with a new polarization vector. In a typical QELSS experiment the incident light is vertically polarized, the first scattering event $\psi$ is measured from the perpendicular to the scattering plane, and the polarization remains perpendicular to the scattering plane, so $\sin(\psi) = 1$. Because in multiple scattering the scattering paths are not confined to a plane perpendicular to the incident polarization axis, in general $\sin(\psi) \neq 1$.  The factor $\sin(\psi)$ arises already in QELSS experiments in which the incident laser polarization lies in the scattering plane.  For example, for HH scattering (as opposed to the common VV experiment) from optically isotropic spheres, at $\theta=90^{o}$ the scattering intensity is zero. 

For small particles, a simple geometric construction relates $\theta$ and $\psi$.  Namely, without loss of generality we may take the scattering event to be at the origin, the incident light to define the $+x$-direction with its polarization defining the $z$-axis, and the scattering vector to be in an arbitrary direction not confined to the $xy$-plane. Defining $(\theta_{s}, \phi_{s})$ to be the polar angles of the scattering direction relative to $\hat{z}$, the $+x$-axis lies at $(\theta_{x}, \phi_{x}) = (\pi/2,0)$ and the angle addition rule gives $\cos(\theta) = \cos(\theta_{s}) \cos(\theta_{x}) - \sin(\theta_{s}) \sin(\theta_{x}) \cos(\phi_{s}-\phi_{x})$, so one has 
\begin{equation}
    q^{2} = 2 k_{0}^{2} (1 - \sin(\theta_{s}) \cos(\phi_{s}) ).
    \label{eq:q2localframe}
\end{equation}
The angle $\theta_{s}$ is not the scattering angle $\theta$ of eq \ref{eq:q1definition}. For the mean-square scattering vector from a single scattering event in a DWS experiment in which all scattering angles are allowed, 
\begin{equation}
    \overline{q^{2}}\equiv (4 \pi)^{-1} \int d \Omega_{s} \sin(\theta_{s})  2 k_{0}^{2} (1 - \sin(\theta_{s}) \cos(\phi_{s}))   = \frac{\pi}{2}  k_{0}^{2}, 
    \label{eq:q2average}
\end{equation}
 The next two moments are $\overline{q^{4}} = \frac{11 \pi}{8}  k_{0}^{4}$ and  $\overline{q^{6}} = \frac{17      \pi}{4}  k_{0}^{6}$.  The corresponding cumulants in a $q^{2}$ expansion are $K_{1} = \pi k_{0}^{2}/2$, $K_{2} \cong -3.083 k_{0}^{4}$ and $K_{3} \cong 0.7473 k_{0}^{6}$. The second cumulant is not negligible with respect to the first, in the sense that the variance $(\mid K_{2}/K_{1}^{2} \mid)^{1/2}$ is $\approx 1.12$. For in-plane scattering in a QELSS experiment, the average of $q^{2}$ over all allowed scattering angles would not be given by eq \ref{eq:q2average}.  It would instead be be proportional to $\int d\theta \sin(\theta) \sin^{2}(\theta/2)$.  The averages for QELSS and DWS differ because for QELSS only in-plane scattering arises, while in DWS out-of-plane scattering events are allowed and important, and because in QELSS with VV scattering the corresponding polarization weighting factor is unity, while in DWS the allowed scattering angles are polarization weighted by $\sin(\theta_{s})$.  A calculation made in the inadequate scalar-wave approximation would overlook this distinction.

\subsection{Fluctuations in the Number of Scattering Events}

In the standard treatment of photon diffusion in a DWS scattering cell, the path length distribution is computed by envisioning photons as random walkers, and solving the diffusion equation as a first-crossing problem to determine the distribution of path lengths.  A pathlength $s$ is approximated as containing precisely $s/\ell^{*}$ steps, the fluctuation in the number of steps arising entirely from differences in the lengths of the various paths.   MacKintosh and John\cite{mackintosh1989a} present an extended treatment for the path length distribution, using a diffusion picture and saddle point methods to establish the mean $\overline{N}$ number of scattering events and the mean-square fluctuation in that number, as averaged over all path lengths.  They treat separately paths involving few scattering events, for which a simple diffusion picture does not accurately yield the distribution of scattering events.

In addition to the fluctuations in $N$ arising from fluctuations in $s$, for a path of given $s$ there are also fluctuations in $N$ that arise because $\ell^{*}$ is only the average length of a path. While a path of length $s$ on the average contains $\overline{N(s)} =s/\ell^{*}$ scattering events, for paths of given physical length $s$ there will also be a fluctuation $\langle (\delta N(s))^{2} \rangle =\overline{N^{2}(s)} - \overline{N(s)}^{2}$ in the number $N(s)$ of scattering events. Scattering is a rate process linear in path length, so it is governed by Poisson statistics.  For paths of fixed length, the mean-square fluctuation is therefore linear in the number of events. 

\subsection{Joint Fluctuation Effect}

For fluctuations within an exponential of a multilinear form, the case here, the error due to neglecting the fluctuation is readily obtained.  Namely, for a function
\begin{equation}
       f(a) = \langle \exp( - a x) \rangle_{x},
       \label{eq:fangeneral}
\end{equation}
where $\langle \cdots \rangle_{x}$ denotes an average over the distribution of $x$, a Taylor series expansion gives
\begin{equation}
      f(a) = \sum_{i=0}^{\infty} \frac{(-a)^{i} \overline{x^{i}}}{i!}.
      \label{eq:fanseries}
\end{equation}
Here $\overline{x^{i}} \equiv  \langle x^{i} \rangle_{x}$ is the $i^{\rm th}$ moment of $x$.  The function $f(a)$ may equally be written
\begin{equation}
     f(a) = \exp\left(\sum_{i=0}^{\infty} \frac{K_{i} a^{i}}{i!}\right).
      \label{eq:fancumulantseries}
\end{equation}
The $K_{i}$ are the cumulants, with $K_{0} =1$ and $K_{1} = \overline{x}$. The higher-order cumulants $K_{i}$, which are the coefficients of $a^{i}$ in a Taylor series expansion for $\exp[ a (x-\overline{x})]$ in powers of $a$, give the effects of the fluctuation.  If $\overline{X} \neq 0$,
\begin{equation}
     K_{2} = \overline{x^{2}} - \overline{x}^{2},
\end{equation}
and
\begin{equation}
     K_{3} = \overline{x^{3}} - 3 \overline{x^{2}} \overline{x} + 2 \overline{x}^{3}.
     \label{eq:cumulantseries}
\end{equation}
An alternative case in which $\overline{X}=0$, and $f(a)$ is a series in $a^{2n}$, is solved as eqs \ref{eq:g1qtaverage}-\ref{eq:displacementexpand2}. 

Cumulant expansions are well behaved, and converge under much the same conditions that Taylor series expansions are convergent.  A cumulant series is particularly interesting if $f(a)$ is very nearly exponential in $a$, because under that condition the relaxation is driven by $K_{1}$ and the higher-order $K_{i}$ are often all  small. Cumulant expansions have already been used implicitly in the above.  For example, the form $\langle \exp(- i {\bf q} \cdot {\bf r}_{i}(t) )\rangle \approx 1 - \frac{ \overline{q^{2}}\ \overline{X^{2}}(t)}{2}$
is the lowest order approximant.

Where do fluctuations (equivalently, higher-order cumulants) modify the field correlation function for diffusing wave spectroscopy?  The variables with interesting fluctuations are the displacement $X$, the mean-square scattering vector $q^{2}$, and the number $N$ of scattering events in a scattering path.  The quantity being averaged is $\langle \exp( - N q^{2} (\Delta X(t)^{2}) \rangle$. Repeated series expansions in $N$, $q^{2}$, and $(\Delta X)^{2}$, through the second cumulant in each variable, using the methods of ref \onlinecite{phillies2005a}, lead to 
\begin{displaymath}
    g^{(1)}_{\rm DWS}(t) = \left\langle \exp\left(- \overline{N} \left[ \frac{\overline{q^{2}}\ \overline{X(t)^{2}}}{2} + 
    \frac{\overline{q^{4}} (\overline{X(t)^{4}} - 3 \overline{X(t)^{2}}^{2})  }{24} + \ldots \right]
     \right.\right.
\end{displaymath}
\begin{equation}
   \left.\left. + \frac{\overline{X(t)^{2}}^{2}}{8} \left[\overline{N}^{2}  (\overline{q^{4}} - \overline{q^{2}}^{2})
               + \overline{q^{2}}^{2}  (\overline{N^{2}} - \overline{N}^{2})\right] + \ldots ) \right)\right\rangle.
    \label{eq:30s}
\end{equation}
Here $\overline{N}$ and $\overline{N^{2}}$ are the average and mean-square number of scattering events for all paths.  One could also take $\overline{N}$ and $\overline{N^{2}}$ to refer to paths of fixed length $s$, with $\langle \cdots \rangle_{s}$ including an average over the path length distribution.  The above equation may be contrasted with the form  $ g^{(1)}_{\rm DWS}(t) = \langle \exp(- \overline{N(s)}\  \overline{q^{2}}\ \overline{X(t)^{2}} ) \rangle_{s}$, eq.\ \ref{eq:DWSspectrum}, obtained by approximating $\overline{N(s)}=s/\ell^{*}$, $\overline{N(s)^{2}} - \overline{N(s)}^{2}=0$, and $\overline{X(t)^{4}} - 3 \overline{X(t)^{2}}^{2}=0$.  

Equation \ref{eq:30s} shows only the opening terms of series in the fluctuations in $X^{2}$, $q^{2}$, and $N$. The first line of eq 30 reflects particle displacements as captured by an individual single-scattering event and iterated $\overline{N(s)}$ times.  The second line reflects the fluctuations from path to path in the total-square scattering vector and the number of scattering events, the fluctuations being $\overline{q^{4}} - \overline{q^{2}}^{2}$ and $\overline{N(s)^{2}} - \overline{N(s)}^{2}$. The time dependence of $g^{(1)}_{\rm DWS}(t)$ in the above arises from the time dependences of $\overline{X(t)^{2}}$ and $\overline{X(t)^{4}}$.  The term $\overline{X^{4}} - 3 \overline{X^{2}}^{2}$ (and terms not displayed of higher order in $X$) reflect the deviation of distribution of particle displacements from a Gaussian. If $q^{2}$ and $N$ were non-fluctuating, the second line of eq 30 would vanish.  Because $q^{2}$ and $N$ do fluctuate,  $g^{(1)}_{\rm DWS}(t)$ gains additional time-dependent terms, not seen in eqs \ref{eq:DWSspectrum} and \ref{eq:displacementexpand2}, but appearing as the second line of eq \ref{eq:30s}.  

On the rhs of \ref{eq:30s}, the lead term of the exponential is the shorter-time approximant $\exp(- \langle N \rangle \langle q^{2}\rangle \overline{X(t)^{2}})$ of Maret and Wolf\cite{maret1987a}.  The basis of the approximation is that the distribution of $N$ has a peak location (approximated as $\overline{N}$), paths having approximately $\overline{N}$ scattering events dominating the distribution, thereby giving the lead term to eq \ref{eq:30s}.  The lead term does not give the true initial slope of $ g^{(1)}_{\rm DWS}(t)$.  Because the decay rate of a path increases with increasing $N$, the minority of paths having particularly large $N$ are responsible for the initial slope of $ g^{(1)}_{\rm DWS}(t)$.  The minority of paths having particularly small $N$ are substantially responsible for the slow decay of $ g^{(1)}_{\rm DWS}(t)$ at long times. Nonetheless, as shown in the following section, there is a sense in which DWS is sensitive to small-displacement particle motions at early times.

\section{Effect of Particle Polydispersity}

As a concrete example of fluctuation effects, we treat the DWS spectrum of a bidisperse system containing two sizes of probe.  The probes perform Langevin-model diffusion in a simple Newtonian solvent. In this very special case, the single-particle field correlation function becomes
\begin{equation}
    g^{(1)}(q, t) = A_{1} \exp(- D_{1} q^{2} t) +  A_{2} \exp(- D_{2} q^{2} t).
    \label{eq:twocomponent}
\end{equation}
Here $A_{1}$ and $A_{2}$ are the scattering cross-sections for the species 1 and 2 particles, respectively. $D_{1}$ and $D_{2}$ are the respective diffusion coefficients. We transform eq \ref{eq:twocomponent} into the canonical form of eq \ref{eq:displacementexpand2} by taking the exponential of the Taylor series of the logarithm of eq \ref{eq:twocomponent}, namely
\begin{equation}
   g^{(1)}(q, t) = (A_{1}+A_{2}) \exp\left( - \frac{(A_{1} D_{1} + A_{2} D_{2})}{(A_{1}+A_{2})}t q^{2}   +  
           \frac{A_{1}A_{2} (D_{1} - D_{2})^{2} }{2 (A_{1}+A_{2})^{2} } t^{2} q^{4}  + \ldots \right).
   \label{eq:twocomponentlogseries} 
\end{equation}
The mean-square displacement is determined by the average diffusion coefficient
\begin{equation}
    \overline{D} = \frac{(A_{1} D_{1} + A_{2} D_{2})}{(A_{1}+A_{2})}.
    \label{eq:dbar}
\end{equation}
The second cumulant of the squared-displacement-distribution is determined by mean-square range of diffusion coefficients
\begin{equation}
    \langle \Delta D^{2} \rangle = \frac{A_{1}A_{2} (D_{1} - D_{2})^{2} }{2 (A_{1}+A_{2})^{2} }.
    \label{eq:dbar2}
\end{equation}
By comparison with eq \ref{eq:30s}, one finds $ \overline{X(t)^{2}}/2 = \overline{D} t$ and $\frac{(\overline{X(t)^{4}} - 3 \overline{X(t)^{2}}^{2})  }{24} =    \langle \Delta D^{2} \rangle
 t^{2}$.  
 
In interpreting $g^{(1)}(q, t)$, the average diffusion coefficient gives the initial linear slope of the spectrum, while on a semilog plot of $g^{(1)}$ against $t$ the $\langle \Delta D^2 \rangle$ term gives a curvature of the spectrum away from its linear slope.  It has been known since the earliest days of the QELSS technique that $g^{(1)}(q, t)$ is quite insensitive to weak polydispersity.  Particles whose diffusion coefficients differ by a factor of two or three only lead to separable modes if the signal-to-noise ratio of the spectrum is extremely high, say, greater than 1000, permitting the spectrum to be followed through three orders of magnitude or more of relaxation. 

For a bidisperse probe mixture the field correlation function for a DWS spectrum becomes
\begin{displaymath}
    g^{(1)}_{\rm DWS}(t) = \left\langle \exp \left(- \overline{N} \left[ \overline{ q^{2}} \ \overline{D} t  +     \langle \Delta D^2 \rangle t^{2}\overline{q^{4}}/2 + \ldots \right] \right. \right.
\end{displaymath}
\begin{equation}
   \left. \left. +  \left[\overline{N}^{2}  (\overline{q^{4}} - \overline{q^{2}}^{2})  \overline{D}^{2} t^{2}/2
               + (\overline{D}\ \overline{q^{2}} t )^{2}(\overline{N^{2}} - \overline{N}^{2})  / 2 \right] + \ldots  \right) \right\rangle.
    \label{eq:bidisperse3}
\end{equation} 

As an example of the implications of eq \ref{eq:bidisperse3}, consider a model bidisperse system similar to the system studied with DWS by Pine, et al.\cite{pine1990a}, which contained mixtures of 198 and 605 nm polystyrene spheres in various concentration ratios.  In convenient reduced units, the model spheres are given diffusion rates $D_{1} \overline{q^{2}} = 1$ and $D_{2} \overline{q^{2}} = 3$.  To simplify the model, the two species contribute equally to scattering so $A_{1} = A_{2}$, leading to $\overline{D} \ \overline{q^{2}} =2$ and $\langle \Delta D^2 \rangle \overline{q^{4}} = 0.5$.   Pine, et al.\cite{pine1990a} do not supply $\overline{N}$ for their experiment, but for cells of dimensions millimeters and mean free paths $\ell^{*}$ of hundreds of microns an interesting representative number might be $N=20$.  If the decay of the field correlation function can be observed over two orders of magnitude, then approximately $\overline{N}  \ \overline{D} \ \overline{ q^{2}} t  \approx 5$, implying $t \approx 0.13$ in reduced units at the largest $t$ observed.  For this $t$, one has  $\langle \Delta D^2 \rangle\overline{q^{4}}  t^{2} /2 \leq 0.01$.  Over the short range of times covered by the experiment, the deviations from single-exponential behavior at a signal-to-noise ratio of 100 is unobservably small .  

In this particular model, the time at which the spectrum has decayed to virtually to zero is so short that the second spectral cumulant $\langle \Delta D^2 \rangle$ has not yet contributed measurably to the spectral relaxation.  The width of the path distribution leads to deviations from a simple exponential at very long times (paths with small $N$)  and short times (paths with large $N$), but over the narrow range of intermediate times at which the DWS spectrum can be obtained, the spectrum is dominated by the first cumulant $\overline{D}\ \overline{ q^{2}}$.  In agreement with these considerations, ref \onlinecite{pine1990a} reports their mixture spectra are consistent with a single relaxation time. This result corresponds to the well-known QELSS result that mixtures of spheres of similar size have QELSS spectra that are very nearly pure exponentials. If the distribution of diffusion coefficients were made adequately wide, the terms in $\langle \Delta D^2 \rangle$ would become observable in the DWS spectrum.  However, the terms in $\langle \Delta D^2 \rangle$ are determined by the mean-fourth power of the particle displacement, so the appearance of these terms in the observable spectrum would mean that the spectrum has ceased to determine the mean-square particle displacement.

If the solution were polydisperse rather than bidisperse, the $X(t)^{2n}$ would be more elaborate.  However, an essentially arbitrary relaxation spectrum can be written
\begin{equation}
    g(t) = \int_{0}^{\infty} A(\Gamma) \exp (- \Gamma t) d \Gamma,
    \label{eq:AGamma}
\end{equation}
where $A(\Gamma)$ is the normalized amplitude for relaxation at $\Gamma$, and as shown by Koppel\cite{koppel1972a} $g(t)$ has a cumulant expansion
\begin{equation}
   g(t) = \exp \left(\sum_{n=0}^{\infty} \frac{K_{n} (-t)^{n}}{n!} \right)
   \label{eq:timecumulants}
\end{equation}
where $K_{1} = \langle \Gamma \rangle \equiv \int d \Gamma A(\Gamma) \Gamma$ is the \emph{time-independent} intensity-weighted average relaxation rate.  Under circumstances numerically similar to the model, the DWS spectrum would only be sensitive to $K_{1}$, because the DWS spectrum decays essentially to zero before $K_{2} =  \langle \Gamma^{2} \rangle - \langle \Gamma \rangle^{2}$ perturbs $g(t)$.  This rapid decay is the dark side of the optical level arm $\overline{N}$ advantage; just as the DWS spectrum is sensitive to very small particle motions, so also it does not readily see motions over larger distances. In contrast, the QELSS spectrum persists out to considerably longer (in natural units $(\overline D \ \overline{q^{2}})^{-1}$) times, so the second cumulant is more readily seen in QELSS spectra.  If $K_{2}$ were substantially larger, it could perturb the DWS spectrum before the DWS spectrum relaxed. 

\section{Effect of Particle Interactions and Other Particle Motion Effects}

As seen above, under reasonable approximations, the phase shift for multiple scattering along a single path factors into a product of single-particle terms
\begin{equation}
   g^{(1)}(q,t) = \left \langle\exp \left(- i {\bf q}_{i} (0) \cdot ( {\bf r}_{i}(t) - {\bf r}_{i}(0) )  \right) \right\rangle
   \label{eq:singleterm}
\end{equation}
each of which refers to the motion of a single particle.  The factorization is permitted because the mean path between scattering events is much larger than the distance over which particle motions are correlated, so that $\langle \delta {\bf r}_{i}(t) \cdot \delta {\bf r}_{i\pm 1}(t)  \rangle = 0$.  While the scattering particles are far apart from each other, their physical neighbors that are not in the same scattering path may perturb their diffusive motions. In understanding the effects of interparticle interactions on QELSS and DWS spectra, it is important to recall\cite{phillies1974a} that two physically distinct diffusion coefficients microscopically characterize the motion of diffusing macromolecules in non-dilute solutions, as studied with light scattering.  One of these, the \emph{mutual} diffusion coefficient, characterizes the relative motion of pairs of diffusing particles and describes the diffusion of particles down a macroscopic concentration gradient.  The other of these, the single-particle or \emph{self} diffusion coefficient, characterizes the motion of single particles through a uniform background.

Comparison with the modern literature on diffusing Brownian particles\cite{phillies1985a,phillies1995a} makes clear that QELSS is routinely applied in two experimental modes that correspond to the two diffusion coefficients.  First, for single scattering from a solution of particles, all of which contribute equally to the scattering, one has for the field correlation function
\begin{equation}
   g_{m}^{(1)}(q,t) = \left \langle \sum_{i=1}^{M} \sum_{j=1}^{M} \exp \left(- i {\bf q} \cdot ( {\bf r}_{i}(t) - {\bf r}_{j}(0) )  \right) \right\rangle
   \label{eq:pairterm}
\end{equation}
Here $i$ and $j$ independently label the $M$ particles in the system. In this mode, QELSS determines the dynamic structure factor $g_{m}^{(1)}(q,t)$ and the mutual diffusion coefficient $D_{m}$ of the diffusing particles.  $D_{m}$ follows from the first cumulant, namely
\begin{equation}
      D_{m} q^{2} = - \lim_{t \rightarrow 0} \frac{\partial g_{m}^{(1)}(q,t)}{\partial t}.
      \label{eq:dmdefinition}
\end{equation}

Second, for scattering from a dilute solution of scattering particles in a perhaps-concentrated solution of non-scattering particles, QELSS determines to good approximation the single-particle correlation function
\begin{equation}
   g_{s}^{(1)}(q,t) =  M^{-1} \left \langle \sum_{i=1}^{M} \exp \left(- i {\bf q} \cdot ( {\bf r}_{i}(t) - {\bf r}_{i}(0) )  \right) \right\rangle.
   \label{eq:singletermform}
\end{equation}
Equation \ref{eq:singletermform} is also the physical basis for optical probe diffusion. The single-particle correlation function that determines the DWS spectrum differs from eq \ref{eq:singletermform} only in that ${\bf q}$ is different for each scattering event.  From $g_{s}^{(1)}(q,t)$ the self diffusion coefficient of the diffusing particles follows as 
\begin{equation}
      D_{s} q^{2} = - \lim_{t \rightarrow 0} \frac{\partial g_{s}^{(1)}(q,t)}{\partial t}.
      \label{eq:dsdefinition}
\end{equation}

In nondilute solutions, the diffusion coefficients $D_{s}$ and $D_{m}$ are both modified by the direct and hydrodynamic interparticle interactions, but not in the same way.  Interparticle interactions also contribute to the higher time cumulants of $g_{m}^{(1)}(q,t)$ and $g_{s}^{(1)}(q,t)$, so for nondilute Brownian particles QELSS spectra are not exponential in time\cite{phillies1995a}.  Correspondingly, for nondilute Brownian particles the displacement distribution function is not a Gaussian, because particle displacements are non-random: Successive displacements of Brownian particles are correlated with each other because interparticle forces have long correlation times.  The assertion that Brownian particles in complex fluids have non-Gaussian displacement distributions has been unambiguously confirmed by the direct experimental measurements of Apgar, Tseng, \emph{et al.}\cite{apgar2000aDp,tseng2001aDp}

The mutual and self diffusion coefficients are usefully written as averages over the hydrodynamic interaction tensor ${\bf b}_{il}$, which describes the retardation in the motion of particle $i$ due to the presence of neighboring particle $l$,  and the interaction tensor ${\bf T}_{ij}$, which describes the motion induced in particle $i$ by a force on particle $j$.  For the drift velocity ${\bf v}_{Di}$ of particle $i$ due to forces on particles $j$, one has\cite{phillies1995a}
\begin{equation}
    {\bf v}_{Di}(t) =  \sum_{j=1}^{N} \mu_{ij} {\bf F}_{j}.
    \label{eq:mobilitydef}
\end{equation}
Here the sum is over all $N$ diffusing particles, $\mu_{ij}$ is the mobility tensor for the $ij$ pair, and ${\bf F}_{j}$ is the force on particle $j$. The mobility tensors are related to the hydrodynamic interaction tensors by
\begin{equation}
    \mu_{ii} = \frac{1}{f_{o}} ({\bf I} + \sum_{\ell \neq i} {\bf b}_{i \ell} + \ldots )
    \label{eq:selfmobility}
\end{equation}
and
\begin{equation}
    \mu_{ij} = \frac{1}{f_{o}} ({\bf T}_{ij} + \ldots )
    \label{eq:pairmobility}
\end{equation}
Here $f_{o}$ is the single particle drag coefficient, ${\bf I}$ is the identity tensor, and in eq \ref{eq:pairmobility} $i \neq j$.  For spherical particles, the hydrodynamic interaction tensors can be written as power series in $a/r_{ij}$, where $a$ is a sphere radius and $r_{ij}$ is the distance between particles $i$ and $j$. In particular,
\begin{equation}
     {\bf T}_{ij} = \frac{3}{4} \frac{a}{r_{ij}}[ {\bf I} + \hat{\bf r}_{ij}\hat{\bf r}_{ij}]
     \label{eq:Toseen}
\end{equation}
is the Oseen tensor approximation to ${\bf T}$, with $\hat{\bf r}_{ij}$ being the unit vector pointing from particle $i$ to particle $j$, while the self-term approximation corresponding to the Oseen tensor is
\begin{equation}
    {\bf b}_{i\ell} = - \frac{15}{4} \left(\frac{a}{r_{i\ell}} \right)^{4} \hat{\bf r}_{i\ell}\hat{\bf r}_{i\ell}.
    \label{eq:bstart}
\end{equation} 

Averages over the hydrodynamic interaction tensors, with due attention to the Brownian motion of the particles, fluctuation-dissipation requirements, and the correct interpretation of the nominal short-time limit of the time derivatives defining the diffusion coefficients give\cite{phillies1985a,phillies1995a}
\begin{displaymath}
      D_{m} q^{2} = \frac{1}{S(k,0)}  \left \langle \frac{k_{B}T}{f_{o}} (q^{2} + \sum_{\ell \neq i =1}^{N} ({\bf q} \cdot {\bf b}_{i\ell} \cdot {\bf q} ) + \exp(i {\bf q} \cdot {\bf r}_{i\ell} ) ({\bf q} \cdot {\bf T}_{i\ell} \cdot {\bf q}) \right.
\end{displaymath}
\begin{equation}
    \left. + \exp(i {\bf q} \cdot {\bf r}_{i\ell} ) i {\bf q} \nabla_{\ell}: ({\bf b}_{i\ell} + {\bf T}_{i\ell }) )
\right \rangle
      \label{eq:dmform}
\end{equation}
and
\begin{equation}
      D_{s} q^{2} =  \left \langle \frac{k_{B}T}{f_{o}} (q^{2} + \sum_{\ell \neq i =1}^{N} \left( {\bf q} \cdot {\bf b}_{i\ell} \cdot {\bf q} ) \right) \right \rangle.
      \label{eq:dsform}
\end{equation}
In the above equations, $S(k,0)$ is the static structure factor, $\nabla_{\ell}$ is taken with respect to the coordinates of particle $\ell$,  and the average is over all possible initial conditions.

To which of these diffusion coefficients is diffusing wave spectroscopy sensitive? DWS responds to the correlation function of eq \ref{eq:singleterm}, which is unmistakeably the single-particle correlation function seen in eq \ref{eq:singletermform}. Correspondingly, DWS is sensitive to the self diffusion coefficient $D_{s}$.  The identification here that DWS measures $D_{s}$ has already been confirmed experimentally: Fraden and Maret\cite{fraden1990a} and Qiu, et al.\cite{qiu1990a} used DWS to measure the diffusion coefficient $\overline{D}$ of polystyrene latex spheres as a function of sphere concentration, finding that the concentration dependence of the $\overline{D}$ measured with DWS agrees with theoretical expectations\cite{beenakker1984a} for $D_{s}$.  

The above treatment of interaction effects differs from some prior analysis.  MacKintosh and John\cite{mackintosh1989a}, their Section III, claim that in nondilute solutions one should replace
eq \ref{eq:singleterm} and their eq 3.6
\begin{equation}
     \langle E(t) E^{*}(0) \rangle \propto  \sum_{n=1}^{\infty} \left \langle \prod_{1}^{n} \langle \int_{{\bf r}{j}}      \epsilon'({\bf r}_{j}, t)
     \epsilon'(0, 0)\rangle_{\rm ens} \exp(i {\bf q}_{j} \cdot {\bf r}_{j}) \right \rangle_{{\bf q}}
     \label{eq:mack0}
\end{equation}
with the rhs of their eq 3.7,
\begin{equation}
     N \mid b(q) \mid^{2} \left \langle \frac{1}{N} \sum{\alpha,\beta} \exp( i {\bf q} \cdot [ {\bf x}_{\alpha}(t) -
{\bf x}_{\beta}(0)] ) \right \rangle.
     \label{eq:mack1}
\end{equation}
In eq \ref{eq:mack0}, the $j$ label $n$ of the scattering particles, and time and space translational invariance have been used to start each particle at the origin at time 0, so ${\bf r}_{j}$ is the displacement of $j$ over time interval $t$.  Eq \ref{eq:mack1} describes a scattering event along some multiple scattering path. $b(q)$ is a scattering cross-section, the sum is over all particles in the system, and ${\bf x}_{\alpha}(t)$ and ${\bf x}_{\beta}(0)$ are the locations of particles $\alpha$ and $\beta$ at times $t$ and $0$.  The sum of particle positions in eq \ref{eq:mack1} is the dynamic structure factor $S(q,t)$.   If this replacement were correct, \emph{which it is not}, then DWS spectra would measure $D_{m}$.  

However, in eq \ref{eq:mack0}, comparison is only made between the position of the same particle at two times.  In contrast, eq \ref{eq:mack1} terms involving the space-time displacements of distinct pairs of particles appear.  A form like eq \ref{eq:mack1} does appear in QELSS theory, in which scattering is coherent, so that the phase relationship between light rays scattered by $\alpha$ and $\beta$ through ${\bf q}$ is determined by the particle positions. Contrary to eq \ref{eq:mack1}: In DWS, the paths leading from the laser to $\alpha$ and $\beta$ are of independent, fluctuating length, so the fields scattered from particles $\alpha$ and $\beta$ through ${\bf q}$ have independent random phases and can not interfere. Furthermore, in DWS a given particle scatters through ${\bf q}$ only if the previous and next particle along the scattering path lie along the pair of rays radiating from $i$ that generate ${\bf q}$.  This constraint is far stronger than the constraint in QELSS, in which each particle scatters light in every direction consistent with the direction of the incident light.  In a liquid, the DWS condition is generally not satisfied, so that even if one considers every multiple scattering path, only some particles scatter through any particular ${\bf q}$.  Contrarily, in eq \ref{eq:mack1} every particle in the system is assumed to scatter light to some other particle through each scattering vector ${\bf q}$.  Therefore, eq \ref{eq:mack1} is not a correct replacement for eq \ref{eq:singleterm}. DWS of non-dilute particles is in the first cumulant sensitive to the $D_{s}$ and not $D_{m}$, in agreement with experimental results of Fraden, et al.\cite{fraden1990a} and Qiu, et al.\cite{qiu1990a}.

Note that eq \ref{eq:bidisperse3} does not require that particle motion be diffusive, only that the particle dynamics be known.  Pine, et al.\cite{pine1990a} discuss experiments on diffusing particles in shear flow, in which the mean-square particle displacement leading to the intermediate-time relaxation of the DWS spectrum includes both diffusive motions (which scale as $q^{2} t^{1}$) and ballistic shear motions (which scale as $q^{2} t^{2}$).  Because the $q$ and $t$ dependences of these motions are \emph{both} known a priori, the mathematical processes used to reach eq \ref{eq:bidisperse3} for diffusive particles lead equally to the DWS spectrum found experimentally for diffusing sheared particles, as shown by Pine, et al.\cite{pine1990a}.

MacKintosh and John\cite{mackintosh1989a} proposes that eq \ref{eq:singletermform} can be written 
\begin{equation}
    g^{(1)}_{s}(q,t) \approx \exp(- q^{2} W(t))
    \label{eq:gaussianform} 
\end{equation}
with $W(t)$ being the time-dependent mean-square particle displacement.  They\cite{mackintosh1989a} cite Hess and Klein\cite{hess1983a} for this approximation, which is mathematically inconsistent with the exact eq \ref{eq:displacementexpand2} in that it is missing the terms in $q^{2n}$, $n > 1$.  Hess and Klein\cite{hess1983a} discuss in detail the differences between $\exp(- q^{2} W(t) )$ and $g^{(1)}_{s}(q,t)$, carefully  emphasizing that they treat $W(t)$  {\em instead} of treating "\ldots the full self-diffusion propagator, which is a complicated function of space and time\ldots ".  Hess and Klein further observe that $g^{(1)}_{s}(q,t)$ is the generating function not only for $W(t)$ but also for the higher-order moments of the particle displacement distribution, those being the higher moments seen in eq \ref{eq:displacementexpand2}.  The use of $W(t)$ may be traced back further to Boon and Yip\cite{boon1991a} who give an expansion equivalent to eq \ref{eq:displacementexpand2}, and note that even though $g^{(1)}_{s}(q,t)$ is not rigorously a Gaussian "it may be a good approximation to treat it as the sum of a Gaussian \emph{and a correction term}". The literature therefore does not support eq \ref{eq:gaussianform} as a correct representation of eq \ref{eq:singletermform}.  

In the discussion above on scattering from polydisperse systems, it was shown that DWS is typically sensitive only to the first cumulant of $g^{(1)}_{s}(q,t)$.  At larger times, the higher cumulants of $g^{(1)}_{s}(q,t)$ are important, but before those times are reached the DWS spectrum may decay to zero.  If the second and higher spectral time cumulants are sufficiently large, the deviation of the spectrum from a single exponential would visible.  However, the higher time cumulants correspond to the higher moments of the displacement distribution function.  If the single-particle function is not a pure exponential, then it does not reflect the mean-square particle displacement.

\section{Analysis}

In this paper, we treat the time dependence of diffusing-wave spectra. We demonstrate that the time dependence of DWS spectra arises, not only from the mean-square particle displacement $\overline{X(t)^{2}}$ , but also from deviations $\overline{X(t)^{4}} -3\overline{X(t)^{2}}^{2}$  from a Gaussian displacement distribution, and also from higher powers $\overline{X(t)^{2}}^{2}$ of the mean-square displacement.  This result does not differ from the corresponding result for QELSS spectra, in which the time dependence of $g^{(1)}_{s}(t)$ arises not only from $\overline{X(t)^{2}}$ but also from higher powers $\overline{X(t)^{2m}}$ of the mean-square displacement. Just as it is erroneous except as a crude approximant to write $\exp(- q^{2} \overline{X(t)^{2}})$ for the general QELSS spectrum, so also it is erroneous to write $\exp(- \langle N \rangle \langle q^{2} \rangle \overline{X(t)^{2}})$ for the general DWS spectrum.   Furthermore, even though fluctuations in $q^{2}$ and $N$ depend only slowly on time, the fluctuations couple to the strongly time-dependent $\overline{X(t)^{2n}}$ and thus to the time dependence of the spectrum at short times.  The analysis of spectra of bidisperse systems shows that the approximant's error can be less serious in DWS spectra; namely, a DWS spectrum decays $\langle N \rangle$ times more rapidly that a QELSS spectrum, so a DWS spectrum may decay to zero before higher time cumulants become significant.

The effect of fluctuations in $N$ and $q^{2}$ on DWS spectra, phrased as deviations from eq \ref{eq:DWSspectrum2}, has been examined by Durian.  Durian\cite{durian1995a} reported Monte Carlo simulations for photons making random walks through a scattering slab.  He computed the path length, number of scattering events, and sum of the squares of the scattering vectors for each path.  These simulations determined fluctuations in the number of scattering events and the total-square scattering vectors, and determined the non-zero effect of these fluctuations on $g^{(1)}_{\rm DWS}(t)$. Durian found that the fluctuation in the total square scattering vector $Y$ increased more slowly than linearly with increasing pathlength $s$.  A slower-than-linear increase is expected for a fluctuating quantity, and does not imply that the second cumulant of $Y$ is negligible for long paths. From Durian's\cite{durian1995a} simulations, an inversion of $g^{(1)}(t)$ via eq \ref{eq:DWSspectrum2} to obtain $\overline{X(t)^{2}}$ has systematic errors, because eq \ref{eq:DWSspectrum2} is inexact.  Fluctuations described here and measured by Durian contribute significantly to the field correlation function. Durian demonstrates circumstances under which fluctuations in $N$ and $q^{2}$ only have effects of some small size on $g^{(1)}(t)$.  He\cite{durian1995a} explains how Monte Carlo simulations could be used to overcome the effect of fluctuations, so as to make determinations of particle motion more accurate than those given by eq \ref{eq:DWSspectrum2}.

Discussions of light scattering spectra are sometimes referred to the Central Limit Theorem. The Central Limit Theorem provides that if a single random variable is constructed as a sum of a large number of {\em identically distributed} subsidiary random variables, then, as the number of subsidiary random variables becomes large, the distribution of the sum variable tends toward a monovariate Gaussian distribution.   This theorem might be applied to describe the distribution of values of the field scattered by a large volume of solution at a single time, or the distribution of changes in the scattered field between any two times.  

However, the utility of the Central Limit Theorem is limited:  

First, the theorem requires that all subsidiary variables be identically distributed.  In many cases of interest, different subsidiary variables in the sum have different distributions. Interesting cases in which the subsidiary variables are not identically distributed include: (1) The scattering Brownian particles have a bidisperse size distribution. In this case, the distribution of particle displacements is different for small and for large particles.  (2) Different particles move in different environments.  For example, the underlying complex fluid is approaching a critical point and has large long-lived local concentration fluctuations, so that particles moving in different fluctuations experience local media having different viscosities.  (3) The underlying complex fluid is viscoelastic, so that the local viscoelastic properties of the fluid are in part determined by the local shear history, i.e., by how far each particle has moved during prior times.  For example, in a viscoelastic fluid, particles that had recently moved a greater-than-average distance might have perturbed the surrounding fluid more than their immobile neighbors would have, so at later times the resistance to their motions might differ from the average resistance, and the distribution of their displacements would differ from the average distribution. 

Second, the Central Limit Theorem gives the distribution of a single variable (which might be the difference between two other variables), but does not guarantee that the joint distribution of three random variables (for example, the values of the scattered field at three times), each pair of which has a joint Gaussian distribution, has an n-variate joint Gaussian distribution.  Doob\cite{doob1942a} shows that to move from the Central Limit Theorem result for two random variables, to the results that three variables are jointly Gaussianly distributed, one needs an additional condition on the evolution of the subsidiary variables, namely that the subsidiary variables are described by a Markoff process.  A simple example of a complex fluid system in which particle motions are not described by a Markoff process is example (3) of the previous paragraph.  If one considers the displacements of particles between a trio of times $t_{1} < t_{2} < t_{3}$, the displacement of particles between any two of those times might have a Gaussian distribution, but the three-fold distribution of particle displacements need not be a trivariate Gaussian, because the distribution of displacements between times $t_{2}$ and $t_{3}$ might have a complicated dependence on the displacement between times $t_{1}$ and $t_{2}$. Note also the recent studies of Lemieux and Durian\cite{lemieux1999a,lemieux2001a} on intermittent dynamic processes leading to non-Gaussian scattering behaviors.

Fortunately, there is an test that determines whether the Central Limit Theorem leads toward a calculation of the spectrum: It is a simple corollary\cite{phillies2005a} of Doob's Theorem\cite{doob1942a} that if the particle displacements have identical Gaussian distributions, and successive displacements are all independent from each other, as is required for the particle motion to correspond to the Langevin model, then the QELSS spectrum is entirely determined by the mean-square particle displacement \emph{and is a pure exponential characterized by a time-independent diffusion coefficient and, correspondingly, a frequency-independent viscosity.} Contrariwise, if the QELSS spectrum is not a pure exponential, the QELSS spectrum is not determined by the mean-square particle displacement.

Even in the special case in which the QELSS spectrum is a simple exponential depending only on $\overline{X(t)^{2}}$, the DWS spectrum does not simplify. From eq 30, even if particle motions are entirely characterized by $\overline{X(t)^{2}}$, so that $\overline{X(t)^{4}} - 3 \overline{X(t)^{2}}^{2}$ and similar higher-order terms all vanish, the fluctuations in $N$ and $q^{2}$ cause the DWS spectrum to depend on $\overline{X(t)^{2}}^{2}$ and higher order terms.  In this special case that probe particles have identical Brownian displacement distributions, it might still be possible to extract $D$ from a DWS spectrum.  However, if the QELSS spectrum were not a simple exponential, for example because the optical probes were bidisperse, then the effect of the $N$ and $q^{2}$ fluctuations is to make it far more challenging to extract the characteristics of the probe motion from a DWS spectrum than from a QELSS spectrum.  The $N$ and $q^{2}$ fluctuations mix fluctuations in the mean-$n^{\rm th}$ displacements, such as $\overline{X(t)^{4}} -\overline{X(t)^{2}}^{2}$, with $\overline{X^{2}}^{2}$ as seen in eq 30.  

\section{When does Diffusing Wave Spectroscopy determine the mean-square particle displacement?}

The underlying issue is interpreting the spectrum of light that has been scattered repeatedly, perhaps many times, by a suspension of optical probe particles diffusing in a simple or complex fluid.  The multiple-scattered spectrum is an elaborate average over sums of uncorrelated single-scattering events.  The best that one can possibly do in interpreting a DWS spectrum is to make a perfect deconvolution of the averages over path length, number of scattering events, and scattering vectors.  A perfect deconvolution would determine from the DWS field correlation function $g_{\rm DWS}^{(1)}(t)$ the single-scattering field correlation function $g_{\rm s}^{(1)}(q, t)$. Even with a perfect deconvolution, the information from DWS spectra can be no better than the information in the single-scattering field correlation function $g_{\rm s}^{(1)}(q, t)$.  We first consider what information is present in the single-scattering spectrum, and then consider additional issues that arise in attempting to deconvolve $g_{\rm DWS}^{(1)}(t)$ to determine $g_{\rm s}^{(1)}(q, t)$.

\subsection{Interpretation of $g_{\rm s}^{(1)}(q, t)$, However Obtained}

As seen from eq \ref{eq:displacementexpand2}, the single-scattering $g_{\rm s}^{(1)}(q, t)$ is determined not only by $\overline{X^{2}}$ but also by all higher moments $\overline{X^{2n}}$, $n > 1$.  The higher moments appear in a variety of combinations, the combinations being non-zero except in the special case that the displacement distribution $P(X)$ is a Gaussian in $X$.  In the special case of a Gaussian $P(X)$, $g_{\rm s}^{(1)}(q, t)$ reduces to a Gaussian in $\overline{X^{2}}$.  Contrariwise, if $P(X)$ is not a Gaussian, then the higher even moments of $X$ all contribute to $g_{\rm s}^{(1)}(q, t)$.

$P(X)$ can be measured directly via video microscopy, at least in systems in which particles do not move too quickly.  For example, Apgar, Tseng, and collaborators\cite{apgar2000aDp,tseng2001aDp}report on mesoscopic probe particles in water:glycerol and in aqueous actin, actin:fascin, and actin:$\alpha$-actinin mixtures. In this work, video light microscopy was used to make repeated measurements of the positions of large numbers of particles at many times; the distribution of particle displacements during various time intervals was computed.  Probe particles in water:glycerol show a Gaussian displacement distribution.  Probe particles in the protein solutions have markedly non-Gaussian displacement distributions, with a displacement distribution far wider (in terms of $X^{2}/\overline{X^{2}}$) and more skewed than observed for particles in water:glycerol.  The direct measurements of Apgar, et al.\cite{apgar2000aDp} and Tseng, et al.\cite{tseng2001aDp} show that any assumption that $P(X)$ for probes in water: actin is a Gaussian is quantitatively incorrect.   Correspondingly, $\overline{X^{2}(t)}$ does not characterize probe motion in these systems.  Analyses of DWS spectra of probe motion in water: actin systems based on eq. \ref{eq:DWSspectrum2} are be invalid, because the underlying assumptions behind eq. \ref{eq:DWSspectrum2} are not satisfied.

Equation \ref{eq:q2r2} for $g_{\rm s}^{(1)}(q, t)$ shows that $P(X)$ and $g_{\rm s}^{(1)}(q, t)$ are linked by a spatial Fourier transform between $X$ and $q$.  If $P(X)$ is a Gaussian in $X$, then its transform $g_{\rm s}^{(1)}(q, t)$ must be the Gaussian $\exp(- q^{2} \overline{X^{2}})$ in $q$ that was assumed in the derivation of eq \ref{eq:DWSspectrum2}.  Contrariwise, if $g_{\rm s}^{(1)}(q, t)$ is not a Gaussian in $q$, then by the same token $P(X)$ is not a Gaussian in $X$, and $g_{\rm s}^{(1)}(q, t)$ depends not only on $\overline{X^{2}}$ but on the higher moments of $X$.  There is a considerable literature on optical probe diffusion as studied with QELSS, often by applying mode decomposition or related spectral analysis methods to $g_{\rm s}^{(1)}(q, t)$.  For probes in HPC solutions, Streletzky, et al.\cite{phillies2003a,streletzky1998a} show: While the relaxation rates of some spectral modes scale as $q^{2}$, for other modes the relaxation rates are not linear in $q^{2}$.  From equation \ref{eq:q2r2}, the component of the displacement distribution $P(X)$ corresponding to a mode that does not relax as $\exp(-a q^{2})$ is necessarily not Gaussian in $X$.  It would be incorrect to infer the viscoelastic properties of systems with  relaxations whose relaxation rates do not scale linearly in $q^{2}$ by using eq \ref{eq:q2r2} or \ref{eq:DWSspectrum2}, because particle motions in these systems would not satisfy the assumptions on which these equations are based.

The relaxation of an arbitrary mode can formally be written $\exp(- t q^{2} D(q,t))$.  With simple diffusion, $D(q,t)$ is a constant having the trivial $q$ and $t$ dependences $q^{0} t^{0}$.  If $D(q,t)$ has nontrivial dependences on $q$ and $t$,  an average over $q$ does not factorize as $\overline{q^{2} D} = \overline{q^{2}} \ \overline{D}$ because $D$ is a function of $q$, contrary to the implicit assumption that the exponent on the rhs of eq \ref{eq:q2r2} was purely quadratic in $q$.

What are the special cases in which eq \ref{eq:DWSspectrum2} provides a correct description of the single-scattering field correlation function?  Two are readily identified. First, the probes might be a monodisperse suspension that diffuses in accord with the Langevin equation, as described by Berne and Pecora\cite{berne1976a}.  Polystyrene latex spheres in water:glycerol afford an example.  The solvent is a simple Newtonian fluid having no viscoelastic memory on observable time scales. In this case, the distribution of particle displacements is Gaussian, and
\begin{equation}
    g^{(1)}_{s}(t) = \exp(- q^{2} D t)
    \label{eq:g1sDt}
\end{equation}
with $D$ a constant. The diagnostic for this case is that a plot of $\log(g^{(1)}_{s}(t))$ is a straight line, starting at the smallest observable time and extending out until the signal fades into the noise.  

Second, the probes might be diffusing in a viscoelastic fluid that has identifiable longest time and distance scales on which relaxation occurs.  Particle motions, over times and distances much longer than the largest relaxation time and distance, satisfy the requirements of the Central Limit Theorem and Doob's Theorem.  Over sufficiently large times and distances, the probes perform simple Brownian motion.  At long times $\log(g^{(1)}_{s}(t))$ against $t$ becomes a straight line, from which a long-time $\overline{X(t)^{2}}$ and $D$ can be extracted.  

To extract a long-time limiting slope from $\log(g^{(1)}_{s}(t))$, a long time linear limit must actually exist, a circumstance that is not guaranteed to arise.  For example, careful QELSS studies show that in some polymer systems $g^{(1)}_{s}(t)$ decays at large $t$ as a stretched exponential in time\cite{phillies2003a}.  These systems have a continuous distribution of relaxation times but not a single longest relaxation time.  In these systems, fitting a straight line to $\log(g^{(1)}_{s}(t))$ at large $t$ is meaningless.  

\subsection{Deconvolution of $g_{\rm DWS}^{(1)}(q, t)$ to Determine $g_{\rm s}^{(1)}(q, t)$}

The reconstruction of the single-scattering field correlation function from the diffusing-wave field-correlation function faces a fundamental challenge. It is fundamentally impossible to
reconstruct a general function of two variables $q$ and $t$, namely $g_{\rm s}^{(1)}(q, t)$ from
a general function of one variable $t$, namely $g_{\rm DWS}^{(1)}(t)$, when the univariate function of $t$ was generated via an average over ${\bf q}$ of the bivariate function.  The issue is simple: information on the $q$-dependence of  $g_{\rm s}^{(1)}(q, t)$ is destroyed by the averaging process. This fundamental limit is mathematical, not physical, and arises from the information-theoretic consequences of taking the average over ${\bf q}$.
 
It might superficially appear that eq \ref{eq:DWSspectrum2} reconstructs $g_{\rm s}^{(1)}(q, t)$  from $g_{\rm DWS}^{(1)}(t)$.  There is no reconstruction here.   Rather, the $q$-dependence of $g_{\rm s}^{(1)}(q, t)$ is taken to be known a priori to be $g_{\rm s}^{(1)}(q, t) \sim \exp(- q^{2} Dt)$, for $D$ independent of $q$. Equation \ref{eq:DWSspectrum2} then only needs to reconstructs the time dependence of $g_{\rm s}^{(1)}(q, t)$, and that only so far as the first time cumulant, which is possible.  The $q$ dependence of $g_{\rm s}^{(1)}(q, t)$ does not need to be diffusive.  Wu, et al.\cite{wu1990a} treat reconstruction for diffusing particles in a shear flow, for which $g_{\rm s}^{(1)}(q, t)$ has known diffusive and shear components.  A $q$-dependence that is known a priori will support reconstruction.  However, without a known form for the $q$-dependence of  $g_{\rm s}^{(1)}(q, t)$,  reconstruction of $g_{\rm s}^{(1)}(q, t)$ from $g_{\rm DWS}^{(1)}(t)$ be made. For probe particles in polymer solutions, a common topic of investigation, the careful work of Streletzky\cite{streletzky1998a,phillies2003a} shows that the $q$ dependence of $g_{\rm s}^{(1)}(q, t)$ is a complicated function of probe radius and polymer concentration that must be determined by systematic experiments. Probe motion in nondilute polymer solutions is therefore fundamentally inaccessible to study by DWS as the method is presently constituted.

\end{document}